\documentclass[a4paper,12pt]{article}

\usepackage{amssymb}
\usepackage{amsmath}
\newcommand{\caviar}{CAViaR}
\usepackage{placeins}
\usepackage{tabularx}
\usepackage{longtable}
\usepackage{adjustbox}
\usepackage{booktabs}
\usepackage{bm}
\usepackage{subfigure}
\usepackage[flushleft]{threeparttable}
\usepackage{wasysym}
\usepackage{latexsym}
\usepackage{color}
\usepackage[round]{natbib}
\usepackage{authblk}

\begin{document}
	\title{Modeling and Forecasting Tail Risk Spillovers: A Component-Based CAViaR Approach}
\author{Demetrio Lacava\footnote{Email: dlacava@unime.it}}
\date{}

\affil[]{Department of Economics, University of Messina, Messina, via dei Verdi, 75, 98122, Italy.}
\maketitle


\begin{abstract}
This paper introduces a new extension of the Conditional Autoregressive Value at Risk (\caviar) model aimed at improving tail risk forecasting across assets. The proposed component-based model, \caviar\ with Spillover Effects (\caviar-SE), decomposes the conditional Value at Risk into a proper-risk component and a spillover component driven by a linear combination of tail risks from influential assets. These assets are selected via a recursive partial correlation algorithm, allowing multiple spillover sources with minimal parameterization. The spillover component acts as a predictable quantile shifter, directly affecting the conditional quantile dynamics rather than the volatility scale. Empirical results on Dow Jones Industrial Average stocks show that spillover effects account for a substantial share of total tail risk and significantly improve out-of-sample tail risk forecasts. Backtesting procedures, together with Model Confidence Set (MCS) analysis, confirm that \caviar-SE provides well-calibrated risk measures and statistically superior forecasts compared to standard and augmented \caviar\ models.
\end{abstract}

	\section{Introduction}

Since the Basel agreements, Value at Risk (VaR) and Expected Shortfall (ES) have been considered standard measures for financial risk: VaR summarizes the maximum loss expected over a given horizon, while ES captures the average loss when this threshold is exceeded. Together, they provide a concise description of downside tail risk, making them leading tools for both risk management and financial regulation. From a risk management perspective, these measures are employed to quantify potential losses over a given period (one day, one week, or one month). From a regulatory viewpoint, VaR and ES represent the basis for the determination of capital requirements aimed at ensuring the stability of the financial system, particularly during periods of market stress. 

The literature has proposed three main approaches for the estimation of VaR and ES. The non-parametric approach, based on historical data without assuming a specific probability distribution: this ensures a high degree of flexibility but makes this approach less suitable for forecasting purposes. Conversely, the parametric approach requires an explicit assumption on the distribution underlying a volatility model, generally a GARCH-type model \citep{Bollerslev:1986}: VaR is then obtained from the corresponding theoretical quantile. While this approach typically provides good forecasting performance, it is highly sensitive to the assumed distributional specification. Finally, the semi-parametric approach can be viewed as an intermediate solution between parametric and non-parametric methods. It relies on a parametric specification for the conditional dynamics, while allowing for a more flexible modeling of the distribution, particularly in the tails. 

Among the semi-parametric models, the Conditional Autoregressive Value at Risk (\caviar) model proposed by \citet{Engle:Manganelli:2004} has proven to be effective in capturing the dynamics of the conditional quantile, with adequate performance both in-sample and out-of-sample. In the \caviar $\,$ framework, the VaR follows an autoregressive structure augmented with an observed variable accounting for the impact of the most recent return on the expected maximum loss. More recently, \citet{Mitrodima:Oberoi:2024} propose an extension of the \caviar\ framework that jointly models VaR and ES through autoregressive tail dynamics, allowing for long-range dependence and horizon-specific aggregation and improving tail-risk forecasting accuracy. 

Generally, VaR is constructed relying exclusively on the returns of the asset under consideration, typically abstracting from the influence of other factors. In this regard, the inclusion of overnight returns \citep{Meng:Taylor:2018} or the High–Low volatility estimator \citep{Chen:Gerlach:Hwang:McAleer:2012} in the VaR process allows to outperform GARCH-based methods and standard \caviar\ models.\footnote{Further evidence on the informative content of intraday range estimators for VaR prediction is also provided by \citet{Fuertes:Olmo:2013} in the GARCH framework.} Furthermore, the inclusion of realized measures of volatility improves the VaR predictive ability relative to baseline specifications, both in the semi-parametric \citep{Vzikevs:Barunik:2015} and in the parametric framework \citep{Brownlees:Gallo:2010}.
However, the \caviar $\,$ models are generally built without accounting for the fact that risk can propagate across assets and markets. It is well known that financial markets are highly integrated, with strong return correlations among stocks. For instance, on May 11, 2012, JPMorgan Chase (JPM) recorded a $-10\%$ return after announcing a massive loss stemming from complex credit derivative positions (the so-called ``London Whale''), while Goldman Sachs (GS) experienced a $-4\%$ decline, despite the absence of firm-specific news, suggesting the presence of downside risk spillovers originating from a single institution. 

In this respect, a limited but growing body of research investigates the propagation of tail risk across markets using tools that explicitly allow for linkages at extreme quantiles. These approaches underscore the importance of modeling tail interdependence directly, beyond second-moment spillover metrics. For example, \citet{GarciaJorcano:SanchisMarco:2022} propose a state-dependent model that captures ES spillovers across different market conditions (tranquil, normal, and volatile). Spillovers in VaR are also analyzed using the so-called Conditional VaR \citep[CoVaR,][]{Adrian:Brunnermeier:2016,Girardi:Ergun:2013}, which measures the risk of the financial system, or of a particular asset, conditional on another institution being in distress. \citet{Xi:Gao::Zhong:2025} combine DCC-GARCH models, network analysis, and quantile regression to investigate how changes in market interconnectedness affect VaR, thereby capturing downside risk spillovers. Finally, \citet{Su:2020} applies quantile variance decomposition to measure extreme risk spillovers by extending the spillover index proposed by \citet{Diebold:Yilmaz:2009} within a quantile regression framework.

In this paper, we analyze potential tail risk spillovers by proposing a new extension of the \caviar\ model of \citet{Engle:Manganelli:2004}, called \caviar\ with Spillover Effects (\caviar-SE). In our specification, the conditional quantile is decomposed into two distinct components capturing, respectively, the asset’s proper risk dynamics and spillover effects. The proper-risk component follows a standard \caviar\ specification, while the spillover component is modeled as an autoregressive process augmented with a proxy for spillover effects. Importantly, spillovers enter the conditional quantile as a predictable quantile shifter, rather than through volatility scaling. This structure allows us to quantify the contribution of spillover-related risk to total tail risk when estimating a \caviar\ specification with latent factors. 
The spillover effect is measured by a linear combination of risk measures associated with a set of influential assets that may act as sources of spillover for the asset under consideration. The set of influential assets is identified using the recursive algorithm proposed by \citet{Gallo:Lacava:Otranto:2025}, which selects the variables exhibiting the highest partial correlation with the asset of interest: this approach accounts for multiple spillover sources while introducing only one additional parameter into the model.

The empirical analysis, based on the assets belonging to the Dow Jones Industrial Average index, reveals several novel findings. First, the identified set of influential assets suggests that spillover effects reflect underlying structural economic mechanisms rather than spurious correlations. Second, the spillover component enters the model with a positive coefficient, indicating that extreme losses in the selected assets are associated with more severe losses for the asset under consideration. Third, the spillover component accounts for the 20\% of total tail risk, at least. Fourth, the proposed specification delivers reliable VaR estimates and successfully passes standard backtesting procedures. Finally, and most importantly, the \caviar-SE model preserves the simplicity and interpretability of the \caviar\ framework, while enhancing its ability to forecast tail risk in interconnected markets, as highlighted by the out-of-sample forecasting analysis.

The paper is structured as follows. Section \ref{sec:model} introduces the theoretical framework, with the proposed model and the selection algorithm presented in Sections \ref{sec:caviar_se}–\ref{sec:selection_algorithm}; Section \ref{sec:dataset} presents the dataset, while Section \ref{sec:empirical_application} discusses the estimation results, with the out-of-sample forecast evaluated in Section \ref{sec:oos}. Finally, Section \ref{sec:conclusion} concludes the paper with some final remarks.

\section{Conditional Autoregressive Value at Risk models}
\label{sec:model}
Let $r_{t}$ be the vector of the returns observed at time $t$ for a given asset
\begin{equation*}
	r_{t}=\sqrt{h_{t}}\eta_{t} \quad \text{with} ~ \eta_{t}\sim i.i.d.\,(0,1).
\end{equation*}
In this standard specification, $h_{t}$ is the conditional variance of returns, while $\eta_{t}$ is the independent and identically distributed innovation term. In this framework, for any given level $\tau$, the $VaR_t(\tau)$ is defined as the threshold that may be exceeded by returns on day $t$, with probability $\tau$. Formally, 

\begin{equation}
	Pr(r_t<VaR_{t}(\tau)|\mathcal{F}_{t-1})=\tau,
\end{equation}
where $\mathcal{F}_{t-1}$ denotes the available information set. In other words, by focusing on the left tail of the distribution, a negative $VaR_t$ is obtained, representing the expectation about the maximum loss in a given period.

It is crucial to correctly estimate the threshold $VaR_t(\tau)$.\footnote{For convenience, we omit the subscript $\tau$ hereafter.} In the semi-parametric approach proposed by \citet{Engle:Manganelli:2004}, the Value at Risk of a given asset follows a dynamic specification that includes an autoregressive component -- capturing the persistence and smooth evolution of the conditional quantile, consistent with the autoregressive behavior typically observed in volatility -- and a term based on an observable variable that measures the impact of the most recent information. This observable term typically corresponds to the asset's returns, with several Conditional Autoregressive Value at Risk (\caviar) specifications available, depending on how returns enter the model. 

When the absolute returns are employed, the resulting specification is the so-called Symmetric Absolute Value (SAV) \caviar, which is usually expressed as

\begin{equation}
	VaR_t=\omega+\beta_1 VaR_{t-1}+\beta_2 |r_{t-1}|,
	\label{eq:sav}
\end{equation}
where $\omega$ is the constant, while $0<\beta_1<1$ captures the persistence. Furthermore, $|r_{t-1}|$ measures the symmetric impact of returns, so that both positive and negative returns have the same effect on the estimated dynamic quantile. This is coherent with volatility dynamics, which can increase following either negative (bad) or positive (good) returns. However, this represents the main limitation of the SAV \caviar. \citet{Engle:Manganelli:2004} also propose the Asymmetric Slope (AS) \caviar, where negative and positive returns enter the model separately. It is specified as

\begin{equation}
	VaR_t=\omega+\beta_1 VaR_{t-1}+\beta_2 r^+_{t-1}+\beta_3 r^-_{t-1},
	\label{eq:as}
\end{equation} 
where $r^+_{t-1}=max(ret_{t-1},0)$ and $r^-_{t-1}=min(ret_{t-1},0)$. In Eq. \ref{eq:as}, $\beta_3$ is expected to be positive, so that a negative return contributes to a decrement in $VaR_t$ (moving away from zero), reflecting the well known increase in risk (i.e., the leverage effect). Similarly, a positive $\beta_2$ leads to an increment in $VaR_t$ (moving towards zero), which translates into an improvement of the maximum expected loss for a given period (i.e., a reduction in absolute terms). Generally, $\beta_3>\beta_2$ testifies the presence of the leverage effect; while $\beta_2$ is expected to be positive or statistically not significant, it may also assume negative values in empirical applications \citep[see, for example, Table 1 in][]{Engle:Manganelli:2004}. Finally, the Indirect GARCH (IG) \caviar, is expressed as in Eq. \ref{eq:ig}
\begin{equation}
	VaR_t=-(\omega+\beta_1 VaR_{t-1}^2+\beta_2 r_{t-1}^2)^{1/2},
	\label{eq:ig}
\end{equation}
where the negative sign is inserted so that $VaR_t$ can still be interpreted as the expectation about the maximum loss in a given period. 

Related to the Value at Risk is the concept of Expected Shortfall (ES), a measure of the
average loss expected in the worst-case scenarios, i.e., beyond a specific VaR, for a given confidence level. In detail, the Expected Shortfall is linked to the \caviar-based VaR through the following positive scaling parametrization
\[
ES_t = [1 + \exp(\gamma)] \, VaR_t,
\]
where $\gamma$ controls the scale of ES relative to the corresponding VaR. Specifically, it determines how much larger the average loss in the tail is compared to the VaR itself. Given the exponential transformation, the strictly positive scaling factor ensures \(ES_t \le VaR_t\).

In this framework, $VaR$ and $ES$ are jointly estimated by minimizing the following negative log-likelihood, which corresponds to the Fissler–Ziegel \citep{Fissler:Ziegel:2016} consistent loss, implemented via an asymmetric Laplace pseudo-likelihood \citep{Taylor:2019}
\begin{equation*}
	\resizebox{1\linewidth}{!}{$
		\mathcal{L}(VaR, ES, r,\Theta) 
		= \sum_{t=1}^{T} \left[ \frac{\mathbf{1}\{r_t \le VaR_{t}\}-\tau}{ES_{t}}\,(VaR_{t} - r_t) + \frac{VaR_{t}}{ES_{t}} + \log(-ES_{t}) - 1 \right].
		$}
\end{equation*}
Here, \(\frac{\mathbf{1}\{r_t \le VaR_{t}\}-\tau}{ES_{t}} (VaR_{t}-r_t)\), captures the quantile violation and penalizes observations below the VaR; \({VaR_{t}}/{ES_{t}}\) and \(\log(-ES_{t})\), normalize the likelihood and ensure the correct scaling; the constant \(-1\) ensures proper alignment with the asymmetric Laplace pseudo-likelihood formulation.

\subsection{\caviar\ with Spillover Effects}
\label{sec:caviar_se}
Based on the idea proposed by \citet{Engle:Lee:1999} in the GARCH framework, \citet{Mitrodima:Oberoi:2024} provide the so-called component \caviar $\,$ specification, which has the merit of distinguishing a short-run component (with the AS, SAV or IG dynamics) from a long-run component that depends on its own lag and on the lagged return of the considered asset. Building upon this idea, we propose a \caviar $\,$ model where the VaR is the sum of two components, representing the proper risk dynamics and the time-varying spillover effect, respectively. By taking the SAV specification as an example, our model, the SAV \caviar $\,$ with Spillover Effects (SAV \caviar--SE), is designed as in Eq. \ref{eq:sav_spill},
\begin{equation}
	\begin{array}{l}
		VaR_{i,t}=q^p_{i,t}+q^s_{i,t}\\[1.5mm]
		q^p_{i,t}=\omega+\beta_1 q^p_{i,t-1}+\beta_2 |r_{t-1}|\\[1.5mm]
		q^s_{i,t}=\varphi_1 q^s_{i,t-1}+ \varphi_2 \mathbf{y}_{t-1}\mathbf{w_i}.
		\label{eq:sav_spill}
	\end{array}
\end{equation}
The proper risk component, $q^p_{i,t}$, follows a standard \caviar $\,$ specification (the SAV, in this case), while the spillover component $q^s_{i,t}$ evolves as an autoregressive process augmented with an exogenous variable measuring the spillover effect, i.e. a linear combination of the risk measure of a set of assets that may have an influence on the tail risk of the \textit{i-th} asset. In other words, in Eq. \ref{eq:sav_spill}, $\mathbf{y}_{t-1}$ is the \textit{(t-1)-th} row of the matrix containing the VaR of all the considered assets and $\mathbf{w_i}$ is the vector of weights constructed according to the selection algorithm proposed by \citet{Gallo:Lacava:Otranto:2025} and described in Section \ref{sec:selection_algorithm}. Following \citet{Engle:Lee:1999}, the model is identified if the long-run component persists more than the short-run component: given the temporary nature of risk spillover, this translates into $\beta_1>\varphi_1$. For the same identifiability reason, the constant is included only in the proper risk component. Furthermore, since $q^s_{i,t}$ is an AR(1), we expect $|\varphi_1|<1$ to ensure a mean-reverting process. It should be noticed that the spillover component does not represent a stochastic innovation to returns. Instead, it acts as a quantile shifter: a predictable, information-driven component that systematically relocates the conditional $\tau$-quantile of the return distribution based on information available at $t-1$. Unlike shocks, quantile shifters adjust the risk boundary ex ante without introducing additional randomness. This allows the component to be interpreted as a systematic part of the conditional Value at Risk, rather than as purely additive noise.
Finally, we expect $\varphi_2>0$, implying that spillover effects from other assets amplify tail risk, increasing the spillover component (in absolute value) and leading to a more negative conditional quantile, thereby resulting in a lower Value at Risk for the \textit{i}-th asset. 

Finally, when $q^p_{i,t}$ follows the AS dynamics, we obtain the AS \caviar--SE
\begin{equation}
	\begin{array}{l}
		VaR_{i,t}=q^p_{i,t}+q^s_{i,t}\\[1.5mm]
		q^p_{i,t}=\omega+\beta_1 VaR_{t-1}+\beta_2 r^+_{t-1}+\beta_3 r^-_{t-1}\\[1.5mm]
		q^s_{i,t}=\varphi_1 q^s_{i,t-1}+ \varphi_2 \mathbf{y}_{t-1}\mathbf{w_i},
		\label{eq:as_spill}
	\end{array}
\end{equation}
while the IG \caviar--SE is defined as
\begin{equation}
	\begin{array}{l}
		VaR_{i,t}=q^p_{i,t}+q^s_{i,t}\\[1.5mm]
		q^p_{i,t}=-(\omega+\beta_1 VaR_{t-1}^2+\beta_2 r_{t-1}^2)^{1/2}\\[1.5mm]
		q^s_{i,t}=\varphi_1 q^s_{i,t-1}+ \varphi_2 \mathbf{y}_{t-1}\mathbf{w_i}.
		\label{eq:ig_spill}
	\end{array}
\end{equation}

Conceptually, our models share common features with the Composite Asymmetric Multiplicative Error Model \citep[ACM,][]{Brownlees:Cipollini:Gallo:2012}, which distinguishes a long-run component from a short-run component of volatility, and the Spillover AMEM \citep[SAMEM,][]{Otranto:2015}, where a spillover component of volatility is included in a Multiplicative Error Model (MEM) for Realized Volatility (RV).

\subsection{Identifying Sources of Risk}
\label{sec:selection_algorithm}
Risk spillover is the transmission of risk from one asset to one other, with higher correlated stocks that are likely to be more influenced each other. Therefore, modeling risk spillovers would require considering all the possible sources of spillover, with a direct implication of an increasing number of parameters. In principle, the resulting over-parametrization issues might be avoided by recurring to suitable variables reduction techniques. 
In this section, we describe a recursive algorithm for the selection of the main sources of spillover (i.e., the set of influential assets) and the construction of the vector of weights. The latter will be used to form a portfolio that summarizes the spillover effect: doing so, one can consider all the possible sources of spillover, for a given asset \textit{i}, at the cost of including only one additional parameter in the model. 

The following selection algorithm \citep{Gallo:Lacava:Otranto:2025} identifies influential assets based on partial correlations: by conditioning on previously selected variables, the algorithm isolates assets that provide incremental information about the tail risk of the asset under consideration, thereby mitigating the impact of multicollinearity and common shocks.

 Let $\bm{X}_t$ be the matrix that contains all the returns series, excluding the $i$-th series. The algorithm consists of the following steps:
\begin{enumerate}
	\item  Compute $Corr( r_{i,t}, {\bm X}_{t-1})$; let $x^*_{t-1}$ be the variable that provides the highest correlation.
	\item  Regress $ r_{i,t}$ on $x^*_{t-1}$, calculate the p-value on its t-stat and the adjusted determination coefficient ${\bar R}^2$. If the p-value$<\alpha$ (the significance level chosen), insert $x^*_{t}$ into the matrix ${\bm Z}_{t-1}$ and all the remaining variables from ${\bm X}_{t-1}$ into the matrix ${\bm V}_{t-1}$. 
	
	\textbf{Stop} otherwise: no spillover effect is detected and the weights ${\bm w_{i}}$ are set to zero. 
	\item  Regress $ r_{i,t}$ on ${\bm Z}_{t-1}$, save the residuals $\xi_{t}$ and compute ${\bar R}^{2}$. 
	
	If ${\bar R}^{2}$ does not improve, \textbf{go to 7}.
	\item  Regress each variable in ${\bm V}_{t-1}$ on ${\bm Z}_{t-1}$ and save the residuals in the matrix ${\bm E}_{t-1}$.
	\item  Compute $Corr(\xi_t,{\bm E}_{t-1})$ and insert $e_{t-1}^*$, i.e., the series with the highest correlation in ${\bm Z}_{t-1}$ and eliminate the corresponding returns series from ${\bm V}_{t-1}$. 
	
	If ${\bm V}_{t-1}$ is empty, \textbf{go to 7}.
	\item  \textbf{Go to 3}.   
	\item Compute the first principal component of the matrix ${\bm X}_{t-1}$ excluding the corresponding variables in ${\bm V}_{t-1}$, and the weights ${\bm w_{i}}$ as its squared loadings (which sum to 1).
\end{enumerate}
The selection of the first influential asset in step 2  depends on the choice of the significance level $\alpha$, which is set equal to 0.10 in the empirical application.
Unlike techniques such as sparse PCA \citep{Zou:Hastie:Tibshirani2006}, our procedure enables the selection of the most important variables while avoiding any arbitrariness associated with the choice of lasso and ridge parameters, and preserving a clear economic interpretation of the selected spillover sources.

\section{The Data Set}
\label{sec:dataset}

\begin{figure}[t!]
	\centering
	\subfigure[]{\includegraphics[height=9cm,width=14cm]{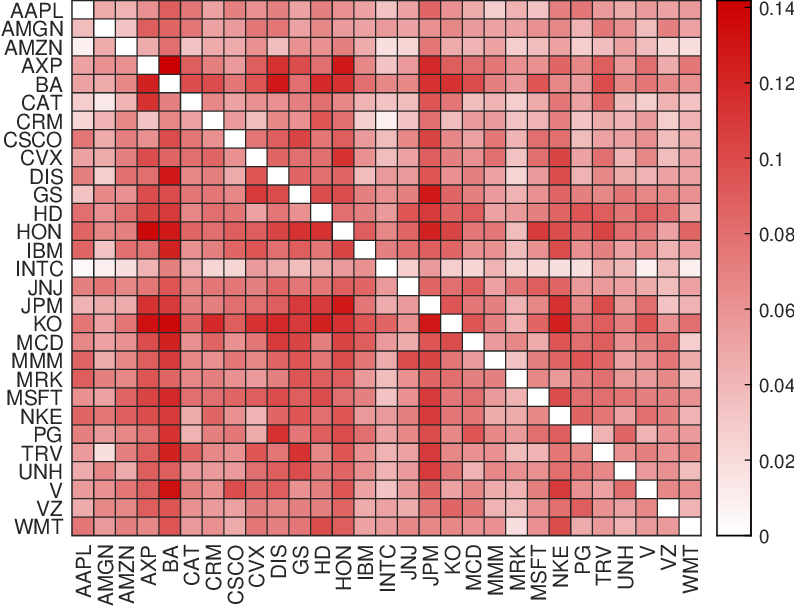}}
	\caption{Heatmap representing the sample cross-quantilogram \citep{Han:Linton:Oka:Whang:2016}. Sample period: March 19, 2008 -- November 14, 2025 }
	\label{fig:correlation_mat}
\end{figure}

The empirical application is based on a daily dataset of the individual stocks included in the Dow Jones Industrial Average\footnote{Due to data availability, we exclude the ticker Dow Inc. The list of the 29 tickers consists of AAPL, AMGN, AMZN, AXP, BA, CAT, CRM, CSCO, CVX, DIS, GS, HD, HON, IBM, INTC, JNJ, JPM, KO, MCD, MMM, MRK, MSFT, NKE, PG, TRV, UNH, V, VZ, and WMT} for the period March 19, 2008 to November 14, 2025. 

A preliminary analysis of quantile dependence can be conducted using the cross-quantilogram \citep{Han:Linton:Oka:Whang:2016}, which measures whether extreme values (high or low, depending on the considered quantile) in one series predict extreme values in another. In practice, it is a measure of the cross-correlation of the quantile-hit process. Figure \ref{fig:correlation_mat} shows the estimated cross-quantilogram for the 0.05 quantile at lag 1. Several interesting findings emerge. First, the correlation is always positive, meaning that when the series in a column crosses its 0.05 quantile at time $t-1$, it is likely that the series in the corresponding row hits its 0.05 quantile at time $t$. In the VaR context, this  would imply that a higher chance of an extreme negative return for asset \textit{j} at time $t-1$ is associated with a higher probability of an extreme negative return for asset \textit{i} at time $t$. Second, contrary to the linear correlation of returns, tail correlation is remarkably lower. Third, for some assets, a moderate tail dependence (above 0.1) is observed, providing a strong justification for including a spillover component in the \caviar\ framework.	

\begin{table}[t]
	\caption{Influential assets for the construction of the vector ${\bm w_{i}}$, identified by applying the recursive algorithm.}\label{tab:source_of_spillover}
	\begin{adjustbox}{max width=1\linewidth,center}
		\begin{tabular}{lccccccccccc}
			\toprule
			Source of spillover & \multicolumn{11}{c}{Tickers} \\
			\midrule
			& AMGN  & CVX  & GS & INTC  & JNJ   & JPM   & MRK   & MSFT  & PG    & TRV   & WMT   \\
			AAPL & 0.071 & 0.079 &       & 0.068 & 0.070 & 0.081 &       & 0.056 & 0.063 & 0.066 & 0.085 \\
			AMGN &       & 0.056 &       & 0.042 & 0.046 & 0.054 &       & 0.040 & 0.045 & 0.046 & 0.061 \\
			AMZN & 0.055 & 0.067 &       & 0.057 & 0.058 & 0.069 &       & 0.044 & 0.050 & 0.054 & 0.068 \\
			AXP  & 0.100 & 0.100 &       & 0.095 & 0.086 & 0.099 &       & 0.077 & 0.088 & 0.082 & 0.114 \\
			BA   & 0.073 & 0.082 &       & 0.070 & 0.068 & 0.082 &       & 0.060 & 0.064 & 0.065 &       \\
			CAT  & 0.085 & 0.093 &       & 0.082 & 0.080 & 0.093 &       & 0.071 & 0.077 & 0.075 & 0.105 \\
			CRM  &       & 0.072 &       & 0.062 & 0.063 & 0.074 &       &       & 0.056 & 0.060 &       \\
			CSCO & 0.084 &       &       & 0.079 & 0.080 &       &       & 0.068 & 0.077 & 0.078 & 0.104 \\
			CVX  & 0.079 &       &       &       &       & 0.085 &       & 0.065 & 0.074 &       & 0.098 \\
			DIS  &       & 0.098 &       & 0.085 &       &       &       & 0.072 &       &       & 0.105 \\
			GS   & 0.095 &       &       & 0.092 &       &       &       & 0.071 &       & 0.076 & 0.109 \\
			HD   & 0.085 & 0.093 & 1.000 & 0.080 & 0.080 & 0.092 &       & 0.069 & 0.077 &       &       \\
			HON  &       & 0.116 &       & 0.098 & 0.099 & 0.114 & 0.282 & 0.088 &       & 0.093 &       \\
			IBM  &       & 0.082 &       &       &       &       &       &       & 0.070 & 0.068 & 0.092 \\
			INTC & 0.065 &       &       &       & 0.064 &       &       & 0.053 &       &       &       \\
			JNJ  &       & 0.063 &       &       &       &       &       &       &       &       &       \\
			JPM  & 0.095 &       &       & 0.090 &       &       &       &       & 0.082 &       &       \\
			KO   &       &       &       &       & 0.053 & 0.063 & 0.224 &       & 0.056 & 0.054 &       \\
			MCD  & 0.063 &       &       &       &       &       &       & 0.052 &       & 0.059 &       \\
			MMM  &       &       &       &       & 0.079 &       &       & 0.070 &       &       &       \\
			MRK  &       &       &       &       &       &       &       &       &       & 0.044 & 0.059 \\
			MSFT &       &       &       &       &       & 0.095 &       &       &       & 0.079 &       \\
			NKE  &       &       &       &       &       &       &       &       &       &       &       \\
			PG   & 0.050 &       &       &       &       &       &       & 0.044 &       &       &       \\
			TRV  &       &       &       &       &       &       & 0.257 &       & 0.073 &       &       \\
			UNH  &       &       &       &       &       &       &       &       &       &       &       \\
			V    &       &       &       &       & 0.076 &       & 0.237 &       &       &       &       \\
			VZ   &       &       &       &       &       &       &       &       & 0.049 &       &       \\
			WMT  &       &       &       &       &       &       &       &       &       &       &       \\
			\bottomrule
		\end{tabular}
	\end{adjustbox}
\end{table}
Table \ref{tab:source_of_spillover} reports the estimated spillover structures across assets, as identified by applying the recursive algorithm outlined in Section \ref{sec:selection_algorithm}, where rows correspond to the assets contributing to the spillover. Overall, the grouping of assets indicates that the identified spillovers reflect structural economic mechanisms rather than spurious correlations. Three main findings emerge. First, assets such as AMGN and MSFT are characterized by a relatively dispersed set of spillover sources, with no single dominant contributor, indicating that spillover-related risk is associated with information originating from multiple segments of the market. Second, financial institutions display distinct spillover structures: while GS exhibits a highly concentrated set of spillover sources, JPM is associated with a more diversified pattern, consistent with broader exposure to different segments of the financial system. Third, assets such as CVX, INTC, and WMT show spillover structures primarily linked to industrial and financial stocks, pointing to a closer association with sources commonly related to aggregate economic conditions.

\section{Empirical Application}
\label{sec:empirical_application}
\begin{table}[t]
	\caption{Estimation results for the SAV, AS and IG \caviar $\,$ models. VaR and ES are jointly estimated by minimizing the \citet{Fissler:Ziegel:2016} loss function. Sample period: March 19, 2008 -- November 14, 2025.}\label{tab:est_baseline}
	\begin{adjustbox}{max width=1\linewidth,center}
		\begin{tabular}{lccccccccccc}
			\toprule
			& AMGN  & CVX  & GS & INTC  & JNJ   & JPM   & MRK   & MSFT  & PG    & TRV   & WMT   \\ \midrule
			&\multicolumn{11}{c}{SAV \caviar-SE}\\
$\omega$  & -0.155 & -0.076 & -0.165 & -0.048 & -0.084 & -0.114 & -0.059 & -0.100 & -0.057 & -0.051 & -0.080 \\
$\beta_1$ & 0.866  & 0.882  & 0.812  & 0.917  & 0.855  & 0.848  & 0.889  & 0.847  & 0.890  & 0.915  & 0.854  \\
$\beta_2$ & -0.148 & -0.197 & -0.279 & -0.145 & -0.197 & -0.229 & -0.175 & -0.239 & -0.170 & -0.129 & -0.210 \\
$\gamma$  & -0.746 & -0.850 & -0.799 & -0.636 & -0.700 & -0.722 & -0.572 & -0.722 & -0.743 & -0.624 & -0.546 \\[2mm]
			\midrule
			&\multicolumn{11}{c}{AS \caviar-SE}\\
$\omega$  & -0.160 & -0.091 & -0.124 & -0.050 & -0.074 & -0.092 & -0.044 & -0.080 & -0.056 & -0.047 & -0.081 \\
$\beta_1$ & 0.861  & 0.886  & 0.861  & 0.916  & 0.874  & 0.867  & 0.918  & 0.874  & 0.899  & 0.923  & 0.854  \\
$\beta_2$ & -0.087 & -0.096 & -0.107 & -0.139 & -0.112 & -0.099 & -0.073 & -0.158 & -0.083 & -0.064 & -0.200 \\
$\beta_3$ & 0.232  & 0.253  & 0.301  & 0.154  & 0.233  & 0.325  & 0.180  & 0.243  & 0.223  & 0.171  & 0.223  \\
$\gamma$  & -0.788 & -0.889 & -0.778 & -0.638 & -0.715 & -0.768 & -0.527 & -0.722 & -0.734 & -0.592 & -0.567 \\[2mm]
			\midrule
			&\multicolumn{11}{c}{IG \caviar-SE}\\
$\omega$  & 0.494  & 0.178  & 0.583  & 0.145  & 0.136  & 0.368  & 0.153  & 0.287  & 0.108  & 0.133  & 0.230  \\
$\beta_1$ & 0.837  & 0.880  & 0.762  & 0.916  & 0.852  & 0.805  & 0.877  & 0.814  & 0.880  & 0.909  & 0.807  \\
$\beta_2$ & 0.188  & 0.267  & 0.481  & 0.174  & 0.243  & 0.366  & 0.214  & 0.384  & 0.233  & 0.152  & 0.294  \\
$\gamma$  & -0.744 & -0.849 & -0.829 & -0.591 & -0.689 & -0.712 & -0.540 & -0.729 & -0.725 & -0.618 & -0.528\\
			\bottomrule
		\end{tabular}
	\end{adjustbox}
\end{table}	
In the empirical application, the proposed models are estimated by setting $\tau = 0.05$. Table \ref{tab:est_baseline} reports the estimation results for the baseline \caviar\ specifications, namely the SAV, AS, and IG models. Overall, the estimated VaR dynamics exhibit a high degree of persistence, with $\hat{\beta}_1$ exceeding 0.75. As expected, in both the SAV and IG models $\hat{\beta}_2$ has the expected sign (i.e., it is negative for the SAV and positive for the IG specification), indicating that the VaR becomes more negative as the asset risk increases, that is, during periods of higher volatility. Consistently, in the AS specification $\hat{\beta}_3>0$, so that the VaR reduces as a response to negative returns. Although a negative estimate of $\hat{\beta}_2$ may seem counterintuitive, the associated worsening in the VaR after positive returns is in line with evidence reported in the field of volatility, which might increase even following positive returns \citep[see, for example, the discussion in the introduction of the SAV model in][]{Engle:Manganelli:2004}. Finally, the $\hat{\gamma}$ coefficient is negative, implying that the scaling factor $1 + \exp(\gamma)$ remains bounded between 1 and 2 and ensures that an ES proportional to, but more conservative than, the corresponding VaR.

\begin{table}[t]
	\caption{Estimation results for the SAV, AS and IG \caviar-SE. VaR and ES are jointly estimated by minimizing the \citet{Fissler:Ziegel:2016} loss function. Sample period: March 19, 2008 -- November 14, 2025.}\label{tab:est_caviarSE}
	\begin{adjustbox}{max width=1\linewidth,center}
		\begin{tabular}{lccccccccccc}
			\toprule
			& AMGN  & CVX  & GS & INTC  & JNJ   & JPM   & MRK   & MSFT  & PG    & TRV   & WMT   \\ \midrule
			&\multicolumn{11}{c}{SAV \caviar-SE}\\
$\omega$    & -0.191 & -0.004 & -0.084 & -0.034 & -0.047 & -0.037 & -0.037 & 0.204  & 0.002  & 0.197  & -0.007 \\
$\beta_1$   & 0.769  & 0.830  & 0.783  & 0.913  & 0.782  & 0.834  & 0.624  & 0.532  & 0.586  & 0.468  & 0.749  \\
$\beta_2$   & -0.127 & -0.143 & -0.244 & -0.134 & -0.176 & -0.203 & -0.233 & -0.246 & -0.173 & -0.215 & -0.175 \\
$\varphi_1$ & 0.000  & 0.000  & 0.000  & 0.000  & 0.000  & 0.000  & 0.000  & 0.000  & 0.000  & 0.000  & 0.000  \\
$\varphi_2$ & 0.325  & 0.583  & 0.429  & 0.154  & 0.275  & 0.340  & 0.644  & 0.899  & 0.510  & 0.818  & 0.437  \\
$\gamma$    & -0.700 & -0.877 & -0.810 & -0.590 & -0.653 & -0.732 & -0.590 & -0.775 & -0.683 & -0.625 & -0.613 \\[2mm]
			\midrule
			&\multicolumn{11}{c}{AS \caviar-SE}\\
$\omega$    & -0.223 & 0.066  & -0.074 & -0.025 & -0.045 & -0.035 & -0.033 & 0.180  & -0.012 & 0.077  & 0.013  \\
$\beta_1$   & 0.710  & 0.727  & 0.863  & 0.921  & 0.683  & 0.840  & 0.590  & 0.508  & 0.549  & 0.790  & 0.763  \\
$\beta_2$   & -0.087 & -0.012 & -0.083 & -0.121 & -0.101 & -0.101 & -0.185 & -0.201 & -0.133 & -0.099 & -0.170 \\
$\beta_3$   & 0.227  & 0.231  & 0.255  & 0.121  & 0.201  & 0.327  & 0.262  & 0.273  & 0.191  & 0.133  & 0.131  \\
$\varphi_1$ & 0.000  & 0.000  & 0.000  & 0.000  & 0.000  & 0.000  & 0.000  & 0.000  & 0.000  & 0.000  & 0.000  \\
$\varphi_2$ & 0.364  & 0.861  & 0.282  & 0.183  & 0.387  & 0.313  & 0.688  & 0.883  & 0.514  & 0.759  & 0.484  \\
$\gamma$    & -0.737 & -0.943 & -0.791 & -0.582 & -0.638 & -0.782 & -0.593 & -0.746 & -0.694 & -0.633 & -0.603 \\[2mm]
			\midrule
			&\multicolumn{11}{c}{IG \caviar-SE}\\
$\omega$    & 0.389  & 0.005  & 0.122  & 0.000  & 0.023  & 0.113  & 0.000  & 0.000  & 0.000  & 0.000  & 0.000  \\
$\beta_1$   & 0.649  & 0.833  & 0.711  & 0.968  & 0.806  & 0.780  & 0.451  & 0.512  & 0.453  & 0.583  & 0.231  \\
$\beta_2$   & 0.117  & 0.088  & 0.262  & 0.037  & 0.104  & 0.249  & 0.108  & 0.129  & 0.053  & 0.081  & 0.081  \\
$\varphi_1$ & 0.000  & 0.000  & 0.000  & 0.000  & 0.000  & 0.000  & 0.000  & 0.000  & 0.000  & 0.000  & 0.000  \\
$\varphi_2$ & 0.358  & 0.556  & 0.508  & 0.321  & 0.290  & 0.300  & 0.743  & 0.680  & 0.518  & 0.619  & 0.553  \\
$\gamma$    & -0.726 & -0.906 & -0.830 & -0.573 & -0.677 & -0.736 & -0.607 & -0.772 & -0.678 & -0.640 & -0.630\\
			\bottomrule
		\end{tabular}
	\end{adjustbox}
\end{table}	

Estimation results for the \caviar $\,$ with spillover effects are presented in Table \ref{tab:est_caviarSE}. As regards the proper risk component, $q^p_{i,t}$ in Eqs. \ref{eq:sav_spill} -- \ref{eq:ig_spill}, the results remain mostly unchanged with respect to the baseline specifications, with high persistence for all the considered assets. Conversely, the spillover component does not display strong persistence ($\hat{\varphi}_1$ is close to zero), suggesting that spillover effects are primarily driven by the exogenous spillover proxy rather than by persistent dynamics. In fact, it is worth noting that $\hat{\beta}_1$ is lower than the corresponding estimates obtained under the baseline specifications. Since the spillover proxy is defined as a linear combination of VaR and exhibits its own dynamic structure, the lower memory of $q^p_{i,t}$ can be interpreted as a reallocation of persistence from the proper to the spillover risk component, with part of the process dynamics being absorbed by the spillover proxy. 

\begin{figure}[t]
	\subfigure[MSFT]{\includegraphics[height=5cm,width=6.8cm]{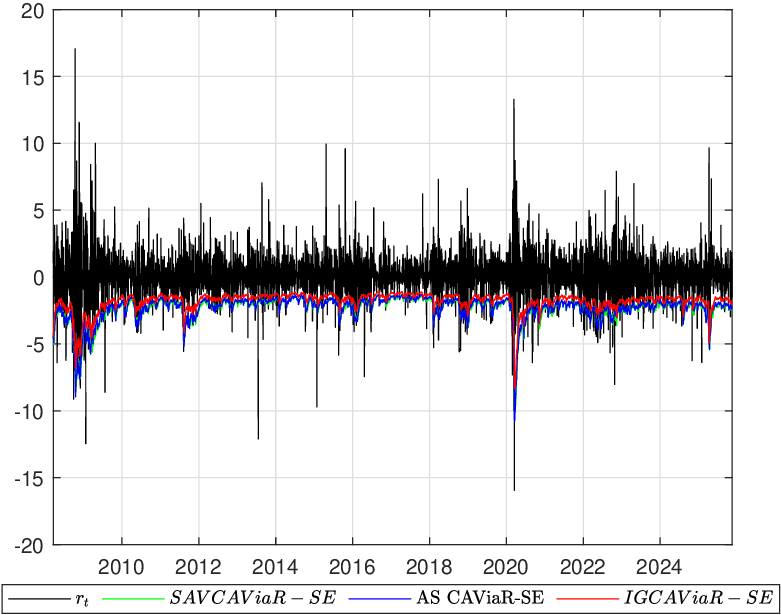}}	\subfigure[JNJ]{\includegraphics[height=5cm,width=6.8cm]{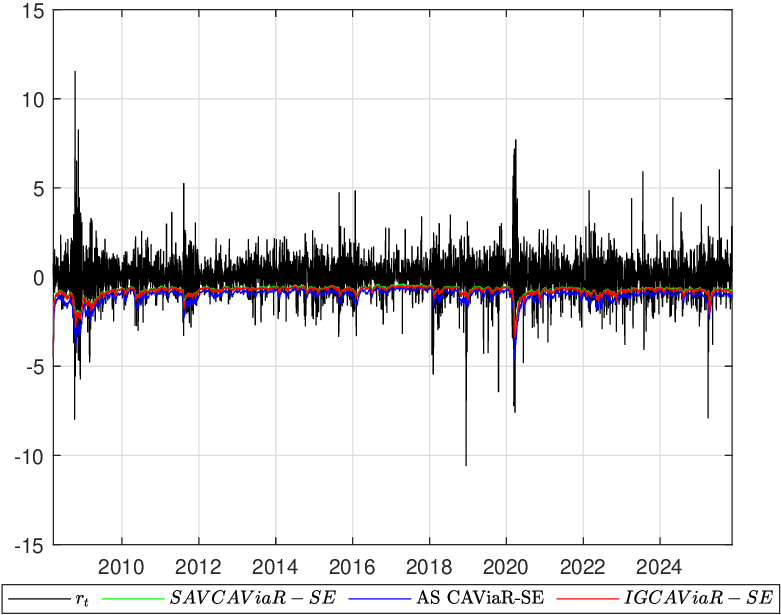}}\\
	\subfigure[GS]{\includegraphics[height=5cm,width=6.8cm]{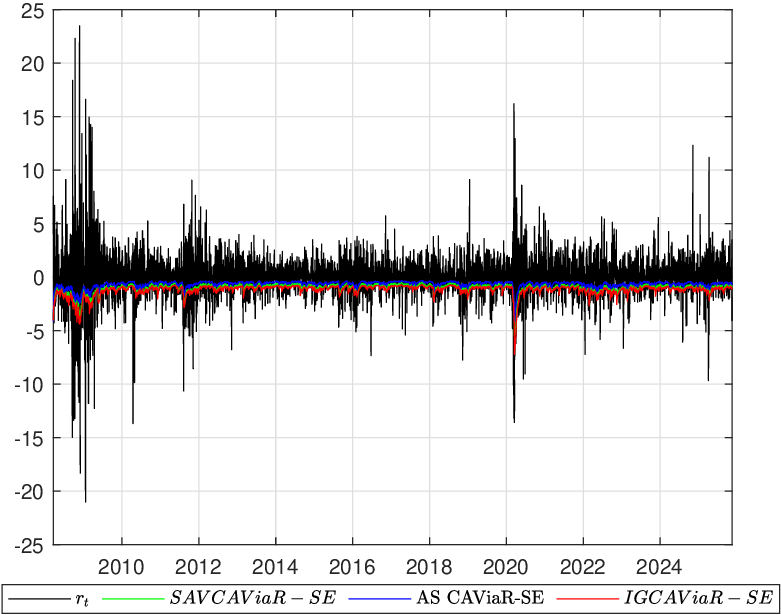}}
	\subfigure[TRV]{\includegraphics[height=5cm,width=6.8cm]{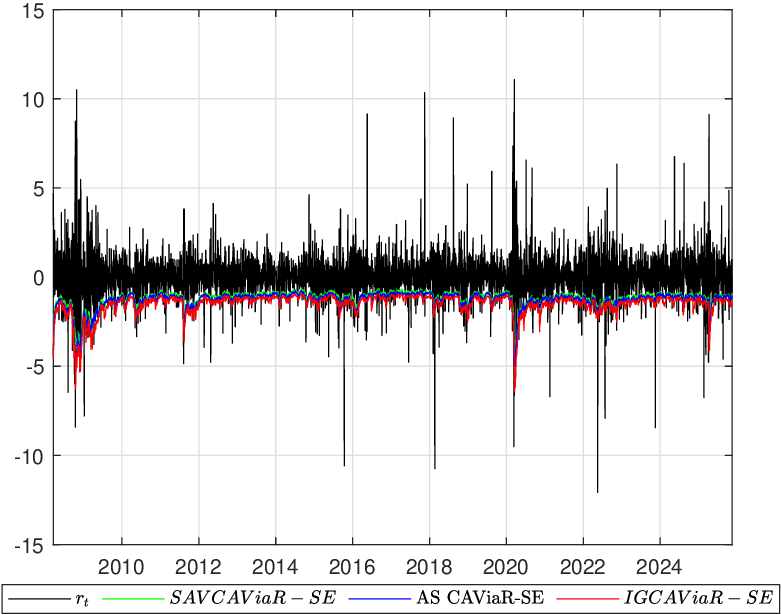}}
	\caption{Spillover component $q^s_{i,t}$ from SAV \caviar-SE (green line), AS \caviar-SE (blue), and IG \caviar-SE (red) models, for selected tickers. Sample period: March 19, 2008 -- November 14, 2025.}
	\label{fig:se_component}
\end{figure}

In line with the primary objective of this analysis -- namely, the identification of spillover effects -- the coefficient of interest is $\hat{\varphi}_2$, which captures the effect of a linear combination of the VaRs of the set of influential assets. The estimated coefficient is consistently positive, suggesting that higher expected losses in the portfolio of influential assets lead to more severe extreme losses for asset \textit{i}. Figure \ref{fig:se_component} displays the evolution of returns (black line) together with the spillover component $q^s_{i,t}$ for representative tickers with different market capitalization, as estimated by the SAV \caviar-SE (green), AS \caviar-SE (blue), and IG \caviar-SE (red) models. The selected tickers are MSFT (belonging to the $1^{\text{st}}$ quantile), JNJ ($2^{\text{nd}}$), GS ($3^{\text{rd}}$), and TRV ($4^{\text{th}}$). The spillover component is negative and exhibits pronounced downward spikes in correspondence with major events, such as the onset of the global financial crisis in late 2008, the COVID-19 pandemic in March 2020, and the growing tensions related to tariffs imposed by the U.S. administration in April 2025. Furthermore, downward spikes are also observed during relatively calm periods, leading to an amplification of tail risk for the \textit{i}-th asset due to shocks that are mainly driven by other assets. 

By taking the ratio ${q^s_{i,t}}/{VaR_{i,t}}$, one can measure the relative contribution of the spillover component to total tail risk. Since both quantities are negative, the ratio is positive and can be interpreted as the share of VaR directly related to the spillover effects. Across models, spillovers account for roughly 20\% of total tail risk for INTC and 80\% for TRV, underscoring the prominent role of the spillover component in shaping the dynamics of extreme risk. This heterogeneity is consistent with differences in the exposure of assets to common risk factors. In particular, for INTC the spillover component plays a relatively limited role compared to the proper-risk component, whereas for TRV tail risk appears to be more closely associated with spillovers originating from other segments of the market.

\begin{figure}[t]
	\subfigure[][\textcolor{black}{\CIRCLE}SAV \caviar-SE \textcolor{red}{\CIRCLE}SAV-X]{\includegraphics[height=5cm,width=6.8cm]{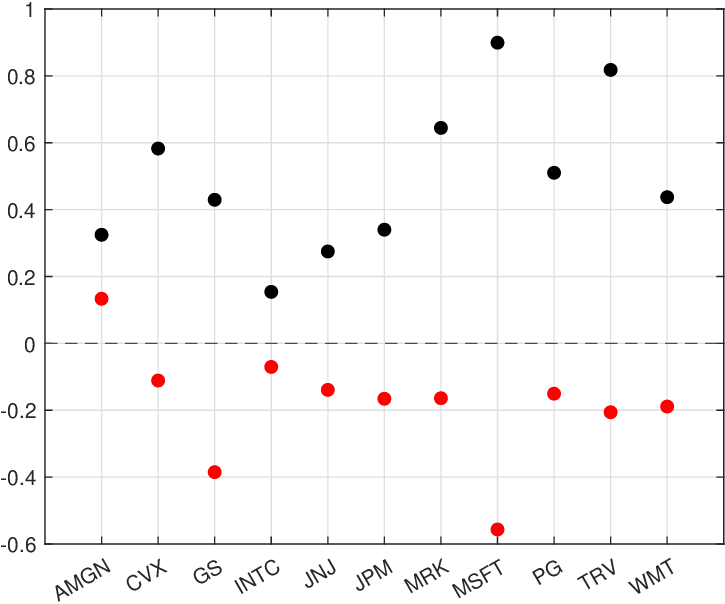}}
	\subfigure[][\textcolor{black}{\CIRCLE}AS \caviar-SE \textcolor{red}{\CIRCLE}AS-X]{\includegraphics[height=5cm,width=6.8cm]{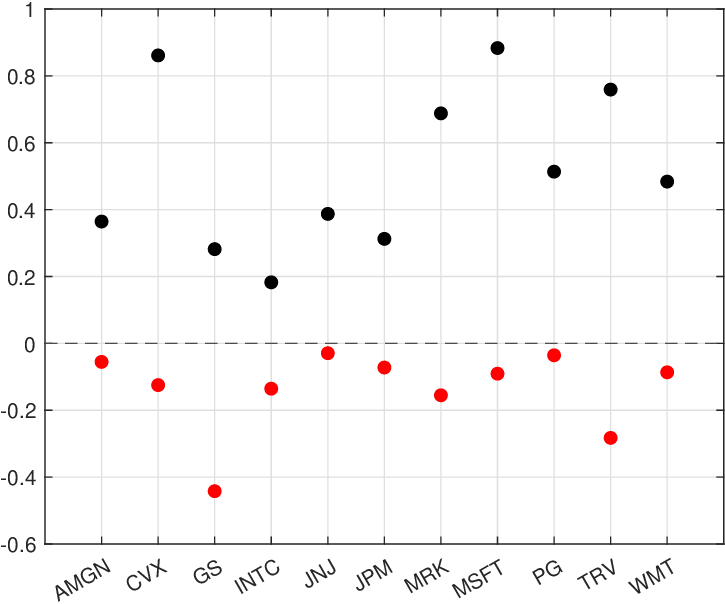}}
	\caption{Spillover effect estimated coefficient from the SAV-X and the AS-X. Sample period: March 19, 2008 -- November 14, 2025.}
	\label{fig:se_vs_x}
\end{figure}

Related to this, Figure \ref{fig:se_vs_x} compares the spillover coefficients obtained from the \caviar-SE model (black dots) with those estimated from the \caviar $\,$ specification augmented with the same spillover proxy included as an exogenous variable (red dots). Since the parameters in the IG specification are constrained to be positive,\footnote{The results (available upon request) show that the spillover coefficient for IG \caviar-SE is consistently higher than for IG-X.} the figure reports the coefficients only for the SAV-X and AS-X models, shown in panels (a) and (b), respectively. With the sole exception of AMGN in the SAV-X specification, the estimated coefficient is negative, so that an increase in the risk of the set of influential assets -- i.e., a more negative value of the spillover proxy -- translates into a higher (less negative) VaR for the \textit{i}-th asset. Given the evidence of positive cross-quantilogram shown in Figure \ref{fig:correlation_mat} and the fact that influential assets are identified based on the highest return correlations, we interpret the negative spillover coefficient as an indication of model misspecification. 

This is further confirmed by the violation rates and the backtests reported in Tables \ref{tab:violation_rate} -- \ref{tab:dq_test}. Specifically, Table \ref{tab:violation_rate} shows that the violation rates are close to the  $\tau=0.05$ level for all \caviar-SE and benchmark specifications. For the \caviar-X models, the violation rate is around 5\% for IG-X, where the spillover coefficient is constrained to be positive, but it exceeds this threshold for the SAV-X and AS-X models. Again, the only exception is AMGN in the SAV-X specification, due to its positive spillover coefficient. Table \ref{tab:uc_test} reports the p-values of the Unconditional Coverage (UC) test \citep{Kupiec:1995}, indicating that the number of VaR violations is inconsistent with the expected frequency: that is, we reject the null for SAV-X and AS-X at the 1\% significance level. Similarly, the Conditional Coverage  test \citep[CC,][]{Christoffersen:1998} in Table \ref{tab:cc_test}  does not reject the null for \caviar $\,$ and \caviar-SE, indicating that VaR violations are consistent with the null hypothesis of independence, while for \caviar-X with a negative spillover coefficient the pattern of violations may show dependence. Finally, the Dynamic Quantile test \citep[DQ,][]{Engle:Manganelli:2004}, as reported in Table \ref{tab:dq_test}, rejects the null of correct quantile prediction for the same models, whereas for \caviar-SE and the benchmark models the p-values exceed the 0.1 threshold. 

\begin{table}[t]
	\caption{Violation Rate for the considered models. Sample period: March 19, 2008 -- November 14, 2025.}\label{tab:violation_rate}
	\begin{adjustbox}{max width=1\linewidth,center}
		\begin{tabular}{lcccccccccc}
			\toprule
			& $SAV$ & $AS$ & $IG$ & SAV-X   & AS-X    & IG-X   & SAV    & AS     & IG  & SAMEM   \\
			& $\caviar-SE$ & $\caviar-SE$ & $\caviar-SE$ &    &     &    &     &      &   &  \\
			\midrule
			AMGN & 5.02\% & 5.02\% & 4.97\% & 5.02\%  & 10.49\% & 5.02\% & 5.00\% & 4.97\% & 4.97\% & 3.76\% \\
			CVX  & 5.00\% & 4.95\% & 5.02\% & 6.71\%  & 10.60\% & 5.00\% & 5.00\% & 5.02\% & 5.00\% & 6.53\% \\
			GS   & 5.00\% & 5.02\% & 5.02\% & 6.23\%  & 8.96\%  & 5.04\% & 5.00\% & 5.00\% & 5.00\% & 5.94\% \\
			INTC & 5.02\% & 5.02\% & 4.97\% & 32.45\% & 17.30\% & 5.00\% & 4.97\% & 5.00\% & 4.97\% & 5.45\% \\
			JNJ  & 5.00\% & 5.06\% & 5.00\% & 8.78\%  & 7.88\%  & 5.02\% & 5.00\% & 5.00\% & 4.97\% & 3.76\% \\
			JPM  & 5.02\% & 4.97\% & 5.02\% & 8.30\%  & 11.21\% & 5.02\% & 5.00\% & 5.00\% & 5.00\% & 5.47\% \\
			MRK  & 5.00\% & 5.02\% & 5.00\% & 12.51\% & 7.22\%  & 5.02\% & 5.06\% & 5.36\% & 5.09\% & 3.67\% \\
			MSFT & 5.00\% & 5.00\% & 5.00\% & 9.97\%  & 13.57\% & 5.02\% & 5.00\% & 5.02\% & 5.02\% & 5.36\% \\
			PG   & 5.02\% & 5.04\% & 5.00\% & 9.95\%  & 11.32\% & 5.02\% & 5.00\% & 5.02\% & 4.97\% & 4.03\% \\
			TRV  & 5.00\% & 4.97\% & 5.00\% & 9.68\%  & 5.60\%  & 5.02\% & 5.00\% & 5.42\% & 4.97\% & 4.73\% \\
			WMT  & 5.02\% & 4.97\% & 5.02\% & 7.09\%  & 12.98\% & 5.02\% & 5.00\% & 5.38\% & 5.00\% & 3.58\% \\
			\bottomrule
		\end{tabular}
	\end{adjustbox}
\end{table}

\begin{table}[t]
	\caption{Unconditional Coverage test \citep{Kupiec:1995} for the considered models. Sample period: March 19, 2008 -- November 14, 2025.}\label{tab:uc_test}
	\begin{adjustbox}{max width=1\linewidth,center}
		\begin{tabular}{lcccccccccc}
			\toprule
			& $SAV$ & $AS$ & $IG$ & SAV-X   & AS-X    & IG-X   & SAV    & AS     & IG & SAMEM    \\
			& $\caviar-SE$ & $\caviar-SE$ & $\caviar-SE$ &    &     &    &     &      &   &   \\
			\midrule
AMGN   & 0.956            & 0.956           & 0.934           & 0.956 & 0.000 & 0.956 & 0.989 & 0.934 & 0.934 & 0.000    \\
CVX    & 0.989            & 0.879           & 0.956           & 0.000 & 0.000 & 0.989 & 0.989 & 0.956 & 0.989 & 0.000    \\
GS     & 0.989            & 0.956           & 0.956           & 0.000 & 0.000 & 0.902 & 0.989 & 0.989 & 0.989 & 0.005    \\
INTC   & 0.956            & 0.956           & 0.934           & 0.000 & 0.000 & 0.989 & 0.934 & 0.989 & 0.934 & 0.179    \\
JNJ    & 0.989            & 0.847           & 0.989           & 0.000 & 0.000 & 0.956 & 0.989 & 0.989 & 0.934 & 0.000    \\
JPM    & 0.956            & 0.934           & 0.956           & 0.000 & 0.000 & 0.956 & 0.989 & 0.989 & 0.989 & 0.158    \\
MRK    & 0.989            & 0.956           & 0.989           & 0.000 & 0.000 & 0.956 & 0.847 & 0.794 & 0.794 & 0.000    \\
MSFT   & 0.989            & 0.989           & 0.989           & 0.000 & 0.000 & 0.956 & 0.989 & 0.956 & 0.956 & 0.282    \\
PG     & 0.956            & 0.902           & 0.989           & 0.000 & 0.000 & 0.956 & 0.989 & 0.956 & 0.934 & 0.002    \\
TRV    & 0.989            & 0.934           & 0.989           & 0.000 & 0.070 & 0.956 & 0.989 & 0.934 & 0.934 & 0.397    \\
WMT    & 0.956            & 0.934           & 0.956           & 0.000 & 0.000 & 0.956 & 0.989 & 0.956 & 0.989 & 0.000    \\
			\bottomrule
		\end{tabular}
	\end{adjustbox}
\end{table}

\begin{table}[t]
	\caption{Conditional Coverage test \citep{Christoffersen:1998} for the considered models. Sample period: March 19, 2008 -- November 14, 2025.}\label{tab:cc_test}
	\begin{adjustbox}{max width=1\linewidth,center}
		\begin{tabular}{lcccccccccc}
			\toprule
			& $SAV$ & $AS$ & $IG$ & SAV-X   & AS-X    & IG-X   & SAV    & AS     & IG   & SAMEM  \\
			& $\caviar-SE$ & $\caviar-SE$ & $\caviar-SE$ &    &     &    &     &      &    &  \\
			\midrule
AMGN   & 0.232            & 0.519           & 0.480           & 0.360 & 0.000 & 0.360 & 0.218 & 0.205 & 0.480 & 0.104    \\
CVX    & 0.130            & 0.941           & 0.360           & 0.000 & 0.000 & 0.675 & 0.342 & 0.857 & 0.500 & 0.006    \\
GS     & 0.675            & 0.969           & 0.857           & 0.000 & 0.000 & 0.714 & 1.000 & 0.961 & 1.000 & 0.005    \\
INTC   & 0.969            & 0.998           & 0.825           & 0.000 & 0.000 & 1.000 & 0.950 & 1.000 & 1.000 & 0.728    \\
JNJ    & 0.841            & 0.538           & 0.940           & 0.000 & 0.000 & 0.930 & 0.841 & 0.940 & 0.950 & 0.790    \\
JPM    & 0.010            & 0.121           & 0.078           & 0.000 & 0.000 & 0.139 & 0.038 & 0.342 & 0.072 & 0.018    \\
MRK    & 0.940            & 0.857           & 0.841           & 0.000 & 0.000 & 0.969 & 0.906 & 0.894 & 0.988 & 0.157    \\
MSFT   & 0.592            & 0.386           & 0.216           & 0.000 & 0.000 & 0.206 & 1.000 & 0.574 & 0.776 & 0.936    \\
PG     & 0.969            & 0.996           & 0.961           & 0.000 & 0.000 & 0.969 & 0.841 & 0.998 & 0.951 & 0.956    \\
TRV    & 0.841            & 0.326           & 0.675           & 0.000 & 0.000 & 0.695 & 0.218 & 0.067 & 0.326 & 0.286    \\
WMT    & 0.998            & 0.951           & 0.930           & 0.000 & 0.000 & 0.930 & 0.961 & 0.969 & 0.961 & 0.637    \\
			\bottomrule
		\end{tabular}
	\end{adjustbox}
\end{table}

\begin{table}[t]
	\caption{Dynamic Conditional Quantile test \citep{Engle:Manganelli:2004} for the considered models. Sample period: March 19, 2008 -- November 14, 2025.}\label{tab:dq_test}
	
	\begin{adjustbox}{max width=1\linewidth,center}
		\begin{tabular}{lcccccccccc}
			\toprule
			& $SAV$ & $AS$ & $IG$ & SAV-X   & AS-X    & IG-X   & SAV    & AS     & IG  & SAMEM   \\
			& $\caviar-SE$ & $\caviar-SE$ & $\caviar-SE$ &    &     &    &     &      &  &    \\
			\midrule
AMGN   & 0.545            & 0.827           & 0.553           & 0.436 & 0.000 & 0.424 & 0.345 & 0.496 & 0.671 & 0.001    \\
CVX    & 0.405            & 0.570           & 0.546           & 0.000 & 0.000 & 0.779 & 0.787 & 0.835 & 0.624 & 0.000    \\
GS     & 0.855            & 0.790           & 0.839           & 0.000 & 0.000 & 0.198 & 0.526 & 0.867 & 0.856 & 0.000    \\
INTC   & 0.732            & 0.742           & 0.484           & 0.000 & 0.000 & 0.255 & 0.630 & 0.841 & 0.503 & 0.431    \\
JNJ    & 0.371            & 0.606           & 0.958           & 0.000 & 0.000 & 0.806 & 0.723 & 0.615 & 0.962 & 0.011    \\
JPM    & 0.011            & 0.280           & 0.081           & 0.000 & 0.000 & 0.213 & 0.019 & 0.633 & 0.054 & 0.010    \\
MRK    & 0.778            & 0.921           & 0.936           & 0.000 & 0.000 & 0.997 & 0.893 & 0.975 & 0.998 & 0.002    \\
MSFT   & 0.916            & 0.863           & 0.568           & 0.000 & 0.000 & 0.713 & 0.974 & 0.814 & 0.896 & 0.428    \\
PG     & 0.747            & 0.984           & 0.838           & 0.000 & 0.000 & 0.199 & 0.177 & 0.848 & 0.681 & 0.039    \\
TRV    & 0.104            & 0.103           & 0.083           & 0.000 & 0.000 & 0.091 & 0.140 & 0.081 & 0.120 & 0.081    \\
WMT    & 0.989            & 0.882           & 0.997           & 0.000 & 0.000 & 0.920 & 0.833 & 0.797 & 0.968 & 0.001 \\
			\bottomrule
		\end{tabular}
	\end{adjustbox}
\end{table}

The last column of Tables \ref{tab:violation_rate} -- \ref{tab:dq_test} reports the backtests for the VaR obtained from the SAMEM model \citep{Otranto:2015},\footnote{Due to data availability, the model is estimated by considering the Garman-Klass volatility estimator \citep{Garman:Klass:1980} rather than the Realized Volatility estimator.} which allows us to assess whether the impact of spillover effects on tail risk is mainly related to the scale of the VaR, rather than to its dynamic quantile component.\footnote{To ensure a fair comparison with the semi-parametric \caviar\ specifications, Value at Risk forecasts from the volatility-based model are obtained assuming Student-t distributed standardized returns. This choice allows for heavy-tailed innovations and avoids imposing overly restrictive distributional assumptions.} Results show that the volatility-based VaR has a violation rate (Table \ref{tab:violation_rate}) statistically different from the nominal level (as indicated by the UC test reported in Table \ref{tab:uc_test}), and it generally fails to correctly capture the dynamics of the conditional quantile (the p-value of the DQ test, Table \ref{tab:dq_test}, is lower than 0.1 in 9 out of 11 cases) even though the violations do not show significant clustering (the null hypothesis under the CC test, in Table \ref{tab:cc_test}, is not rejected for most of the considered assets).

In conclusion, the evidence from the backtesting procedures points to a misspecification of the \caviar-X models when the spillover effect is estimated to be negative. In these cases, the implied VaR dynamics are economically counterintuitive and fail to satisfy standard coverage and independence properties. By contrast, the \caviar-SE specification delivers well-calibrated VaR estimates and consistently passes all backtesting tests. Finally, disentangling the spillover component from the proper-risk component leads to improved tail risk estimation relative to volatility-based approaches, indicating that cross-market interactions may influence tail behavior beyond their impact on second moments. Taken together, these findings provide evidence that an explicit spillover component is required in the modeling of tail risk dynamics, even though the estimated autoregressive coefficient $\hat{\varphi}_1$ is close to zero.

\subsection{Out-of-sample Evaluation}
\label{sec:oos}	

The forecasting evaluation of the proposed models is conducted using a rolling-window scheme. Specifically, the models are estimated over a fixed window of 3,974 observations with a step size of one. At each step, both VaR and ES are forecasted, yielding a total of 470 one-step-ahead out-of-sample point forecasts. The comparison is based on the Model Confidence Set \citep[MCS,][]{Hansen:Lunde:Nason:2011}, a statistical procedure that identifies the set of models with the highest predictive performance, in terms of a given loss function and at a given confidence level. In detail, VaR and ES are jointly evaluated through the FZ0 loss proposed by \citet{Fissler:Ziegel:2016}, with the normalization suggested by \citet{Patton:Ziegel:Chen:2019} that fixes the scale of the loss and facilitates robust empirical comparison of VaR–ES forecasts. It is formally defined as
\small{$$ FZ0_{T+1}=\frac{1}{\tau \widehat{ES}_{T+1}} 1_{r_{T+1}\leq \widehat{VaR}_{T+1}}(\widehat{VaR}_{T+1}-r_{T+1})+\frac{\widehat{VaR}_{T+1}}{\widehat{ES}_{T+1}}+log(-\widehat{ES}_{T+1})-1,$$} 

\noindent where $\widehat{VaR}_{T+1}$ and $\widehat{ES}_{T+1}$ denote the one-step-ahead forecasts based on the  information set available up to time \textit{T}.
Table \ref{tab:mcs_oos} reports the MCS p-values related to the range statistic $T_R$. Models belonging to the confidence set at the 0.25 significance level are highlighted in bold. Several results emerge. First, the proposed \caviar-SE specification delivers strong and robust forecasting performance, remaining in the superior set of models for the majority of assets and being excluded only in a limited number of cases. Second, augmented \caviar-X models exhibit more heterogeneous performance, with the SAV-X and AS-X variants occasionally included in the confidence set. In particular, for JPM and MSFT, the AS-X specification is identified as the preferred model ($p-value=1$) despite displaying negative spillover coefficient in the full-sample analysis. This result can be explained by noting that, within the rolling-window framework, the spillover coefficient turns positive in approximately 40\% (JPM) and 30\% (MSFT) of the estimation windows, thereby contributing to improved tail risk forecasts and superior out-of-sample performance. Third, among baseline \caviar\ models, the AS specification is occasionally preferred, while the SAV and IG are rejected for most assets. Finally, incorporating the spillover effect directly into the \caviar\ specification, rather than in the volatility dynamics, improves the VaR–ES forecasts, as the SAMEM model is consistently excluded from the superior set of models (last column in Table \ref{tab:mcs_oos}). Overall, these findings indicate that the \caviar-SE model provides reliable out-of-sample VaR--ES forecasts, and that both the sign and the stability of the spillover coefficient -- as well as its direct inclusion in the \caviar\ specification -- play a crucial role in determining the predictive accuracy of augmented \caviar\ models.

\begin{table}[t]
	\caption{P-values from the Model Confidence Set (MCS) based on the FZ0 loss function for one-step-ahead out-of-sample forecasts. Models belonging to the confidence set at the 0.25 significance level are highlighted in bold.}\label{tab:mcs_oos}
	\begin{adjustbox}{max width=1\linewidth,center}
		\begin{tabular}{lcccccccccc}
			\toprule
			& $SAV$ & $AS$ & $IG$ & SAV-X   & AS-X    & IG-X   & SAV    & AS     & IG  & SAMEM   \\
			& $\caviar-SE$ & $\caviar-SE$ & $\caviar-SE$ &    &     &    &     &      &  &    \\
			\midrule
AMGN & \textbf{0.887} & \textbf{0.887} & 0.012          & 0.012          & \textbf{0.952} & 0.012 & \textbf{0.952} & \textbf{1.000} & \textbf{0.887} & 0.000 \\
CVX  & 0.009          & 0.074          & 0.001          & 0.001          & 0.000          & 0.000 & 0.212          & \textbf{1.000} & 0.212          & 0.152 \\
GS   & 0.012          & 0.012          & 0.012          & 0.000          & 0.012          & 0.000 & 0.012          & \textbf{1.000} & 0.012          & 0.001 \\
INTC & 0.000          & 0.000          & \textbf{1.000} & 0.000          & 0.000          & 0.000 & 0.000          & 0.000          & 0.000          & 0.000 \\
JNJ  & \textbf{0.370} & \textbf{1.000} & 0.073          & \textbf{0.370} & 0.073          & 0.009 & \textbf{0.370} & \textbf{0.370} & \textbf{0.370} & 0.000 \\
JPM  & 0.003          & 0.006          & 0.006          & \textbf{0.816} & \textbf{1.000} & 0.000 & 0.006          & \textbf{0.816} & 0.006          & 0.000 \\
MRK  & \textbf{1.000} & \textbf{0.996} & \textbf{0.278} & 0.000          & 0.000          & 0.000 & 0.000          & 0.000          & 0.000          & 0.000 \\
MSFT & 0.000          & 0.000          & 0.000          & 0.000          & \textbf{1.000} & 0.000 & 0.247          & \textbf{0.620} & 0.247          & 0.179 \\
PG   & \textbf{0.957} & \textbf{1.000} & \textbf{0.716} & 0.051          & 0.000          & 0.015 & 0.051          & 0.149          & 0.204          & 0.000 \\
TRV  & \textbf{1.000} & \textbf{0.639} & 0.244          & 0.046          & 0.244          & 0.046 & 0.244          & 0.244          & 0.046          & 0.000 \\
WMT  & \textbf{0.317} & \textbf{0.932} & \textbf{1.000} & 0.004          & \textbf{0.317} & 0.004 & 0.004          & 0.024          & 0.004          & 0.000\\
			\bottomrule
		\end{tabular}
	\end{adjustbox}
\end{table}

\section{Final remarks}
\label{sec:conclusion}
In this paper, we propose a new class of models to analyze how spillover effects impact the dynamics of VaR and ES. In particular, we suggest variants of the standard \caviar\ models, called \caviar\ with Spillover Effects (\caviar-SE) in which the conditional quantile is the sum of two components, one capturing the contribution of spillover effects, the other reproducing the traditional dynamics of tail risk. The spillover effect is proxied through a linear combination of the VaR of a set of influential assets (identified via a recursive selection algorithm) that may act as sources of spillover for the asset under consideration. Moreover, the component-based structure of the model allows us to isolate and extract a distinct signal associated with spillover-driven tail risk.

The empirical application, based on individual stocks from the Dow Jones Industrial Average index, provides evidence of superior performance of the \caviar-SE relative to both the standard \caviar\ and the \caviar-X, in which the same spillover proxy enters the model as an exogenous regressor. In particular, the results highlight that the definition and modeling of the spillover component are crucial for correctly capturing spillover effects and for obtaining well-calibrated VaR estimates that successfully pass standard backtesting procedures. Furthermore, we find that spillover effects account for a substantial fraction of tail risk, as the share of total risk directly attributable to the spillover component is at least 20\%. 

Crucially, the inclusion of the spillover component also improves out-of-sample tail risk forecasting performance, as documented by the Model Confidence Set (MCS) results. Finally, the \caviar-SE is systematically preferred when compared with a parametric model in which spillover effects enter the conditional volatility dynamics, suggesting that spillovers primarily operate through the conditional quantile dynamics rather than through the scale of the VaR.

\section*{Conflict of interest disclosure} 
\noindent No competing interest is declared.


  \bibliographystyle{elsarticle-harv} 
  \bibliography{biblio}

\end{document}